\documentclass[12pt]{iopart}
\usepackage{graphicx}

\def\beq{\begin{equation}}
\def\eeq{\end{equation}}
\def\ber{\begin{eqnarray}}
\def\eer{\end{eqnarray}}
\def\l{\Lambda}
\def\lsim{\
  \lower-1.5pt\vbox{\hbox{\rlap{$<$}\lower5.3pt\vbox{\hbox{$\sim$}}}}\ }
\def\gsim{\
  \lower-1.5pt\vbox{\hbox{\rlap{$>$}\lower5.3pt\vbox{\hbox{$\sim$}}}}\ }

\def\etal{{\it et al}}

\def\b{{\rm b}}
\def\m{{\rm m}}

\def \lleq {\
  \lower-1.5pt\vbox{\hbox{\rlap{$<$}\lower5.3pt\vbox{\hbox{$\sim$}}}}\ }
\def \ggeq {\
    \lower-1.5pt\vbox{\hbox{\rlap{$>$}\lower5.3pt\vbox{\hbox{$\sim$}}}}\ }

\def\lsim{\
  \lower-1.5pt\vbox{\hbox{\rlap{$<$}\lower5.3pt\vbox{\hbox{$\sim$}}}}\ }
  \def\gsim{\
    \lower-1.5pt\vbox{\hbox{\rlap{$>$}\lower5.3pt\vbox{\hbox{$\sim$}}}}\ }

\begin{document}

\title[Cosmic Mimicry: Is LCDM a Braneworld in Disguise\,?]
{Cosmic Mimicry: Is LCDM a Braneworld in Disguise\,?}

\author{Varun Sahni\dag,\ Yuri Shtanov\ddag\ and Alexander Viznyuk\ddag}

\address{\dag\ Inter-University Centre for Astronomy and Astrophysics, Post Bag 4,
Ganeshkhind, Pune 411~007, India}

\address{\ddag\ Bogolyubov Institute for Theoretical Physics, Kiev 03143, Ukraine}

\eads{\mailto{varun@iucaa.ernet.in}, \mailto{shtanov@bitp.kiev.ua},
\mailto{viznyuk@bitp.kiev.ua}}

\begin{abstract}
For a broad range of parameter values, braneworld models display a remarkable
property which we call {\em cosmic mimicry\/}. Cosmic mimicry is characterized
by the fact that, at low redshifts, the Hubble parameter in the braneworld
model is virtually indistinguishable from that in the LCDM ($\Lambda$ + Cold
Dark Matter) cosmology. An important point to note is that the $\Omega_{\rm m}$
parameters in the braneworld model and in the LCDM cosmology can nevertheless
be quite different. Thus, at high redshifts (early times), the braneworld
asymptotically expands like a matter-dominated universe with the value of
$\Omega_{\rm m}$ inferred from the observations of the local matter density. At
low redshifts (late times), the braneworld model behaves almost exactly like
the LCDM model but with a {\em renormalized\/} value of the cosmological
density parameter $\Omega_{\rm m}^{\rm LCDM}$. The value of $\Omega_{\rm
m}^{\rm LCDM}$ is smaller (larger) than $\Omega_{\rm m}$ in the braneworld
model with positive (negative) brane tension. The redshift which characterizes
cosmic mimicry is related to the parameters in the higher-dimensional
braneworld Lagrangian. Cosmic mimicry is a natural consequence of the
scale-dependence of gravity in braneworld models. The change in the value of
the cosmological density parameter (from $\Omega_{\rm m}$ at high $z$ to
$\Omega_{\rm m}^{\rm LCDM}$ at low $z$) is shown to be related to the spatial
dependence of the effective gravitational constant $G_{\rm eff}$ in braneworld
theory. A subclass of mimicry models lead to an older age of the universe and
also predict a redshift of reionization which is lower than $z_{\rm reion}
\simeq 17$ in the LCDM cosmology. These models might therefore provide a
background cosmology which is in better agreement both with the observed quasar
abundance at $z \gsim 4$ and with the large optical depth to reionization
measured by the Wilkinson Microwave Anisotropy Probe.
\end{abstract}

\pacs{04.50.+h, 98.80.Hw}

\maketitle

\section{Introduction}

Braneworld models have been extensively applied to cosmology, where they
demonstrate qualitatively new and very interesting properties (see
\cite{maartens04,sahni05} for recent reviews).  Theories with the simplest
generic action involving scalar-curvature terms both in the bulk and on the
brane are not only good in modeling cosmological dark energy but, in doing so,
they also exhibit some interesting specific features, for example, the
possibility of superacceleration (supernegative effective equation of state of
dark energy $w_{\rm eff} \leq -1$) \cite{SS,ls} and the possibility of
cosmological loitering even in a spatially flat universe \cite{loiter}.

It was noted some time ago \cite{MMT} that the cosmological evolution in
braneworld theory, from the viewpoint of the Friedmann universe, proceeds with
a time-dependent gravitational constant. In this letter, we further study this
property and report another interesting feature of braneworld cosmology, which
we call ``cosmic mimicry.'' It turns out that, for a broad range of parameter
values, the braneworld model behaves {\em exactly as a LCDM ($\Lambda$ + Cold
Dark Matter) universe\/} with different values of the effective cosmological
density parameter $\Omega_{\rm m}$ at different epochs.
This allows the model to share many of the attractive features of LCDM
\cite{ss00,sahni04} but with the important new property that the
cosmological density parameter inferred from the observations of the large-sale
structure and cosmic microwave background (CMB) and that determined from
neoclassical cosmological tests such as observations of supernovae (SN) can
potentially have different values.

An important feature of this model is that, although it is very similar to LCDM
at the present epoch, its departure from ``concordance cosmology'' can be
significant at intermediate redshifts, leading to new possibilities for the
universe at the end of the ``dark ages'' which may be worth exploring.

We relate the ``mimicry'' properties of the braneworld cosmology with the
properties of gravity in braneworld theories.  In particular, we show that the
change in the cosmological density parameter $\Omega_{\rm m}$ as the universe
evolves can be related to the spatial scale dependence of the effective
gravitational constant in braneworld theory \cite{vdvz}. This can have
important consequences for cosmological models based on the braneworld theory
and calls for more extensive analysis of their cosmological history.

\section{Cosmic mimicry} \label{mimicry}

We consider the simplest generic braneworld model with an action of the form
\begin{equation} \label{action}
\fl S =  M^3 \left[\int_{\rm bulk} \left( {\cal R} - 2 \Lambda_{\rm b} \right)
- 2 \int_{\rm brane} K \right] + \int_{\rm brane} \left( m^2 R - 2 \sigma
\right) + \int_{\rm brane} L \left( h_{ab}, \phi \right) \, .
\end{equation}
Here, ${\cal R}$ is the scalar curvature of the metric $g_{ab}$ in the
five-dimensional bulk, and $R$ is the scalar curvature of the induced metric
$h_{ab} = g_{ab} - n_a n_b$ on the brane, where $n^a$ is the vector field of
the inner unit normal to the brane, which is assumed to be a boundary of the
bulk space, and the notation and conventions of \cite{Wald} are used. The
quantity $K = h^{ab} K_{ab}$ is the trace of the symmetric tensor of extrinsic
curvature $K_{ab} = h^c{}_a \nabla_c n_b$ of the brane. The symbol $L (h_{ab},
\phi)$ denotes the Lagrangian density of the four-dimensional matter fields
$\phi$ whose dynamics is restricted to the brane so that they interact only
with the induced metric $h_{ab}$. All integrations over the bulk and brane are
taken with the corresponding natural volume elements. The symbols $M$ and $m$
denote the five-dimensional and four-dimensional Planck masses, respectively,
$\Lambda_{\rm b}$ is the bulk cosmological constant, and $\sigma$ is the brane
tension.

Action (\ref{action}) leads to the Einstein equation with cosmological constant
in the bulk:
\begin{equation} \label{bulk}
{\cal G}_{ab} + \Lambda_{\rm b} g_{ab} = 0 \, ,
\end{equation}
while the field equation on the brane is
\begin{equation} \label{brane}
m^2 G_{ab} + \sigma h_{ab} = \tau_{ab} + M^3 \left(K_{ab} - h_{ab} K \right) \,
,
\end{equation}
where $\tau_{ab}$ is the stress--energy tensor on the brane stemming from the
last term in action (\ref{action}).

The cosmological evolution on the brane that follows from (\ref{bulk}) and
(\ref{brane}) is described by the main equation \cite{SS,CH,Shtanov,Deffayet}
\begin{equation} \label{solution}
H^2 + {\kappa \over a^2} = {\rho + \sigma \over 3 m^2} + {2 \over \ell^2}
\left[1 \pm \sqrt{1 + \ell^2 \left({\rho + \sigma \over 3 m^2} - {\Lambda_{\rm
b} \over 6} - {C \over a^4} \right)} \right] \, ,
\end{equation}
where  $C$ is the integration constant describing the so-called ``dark
radiation'' and corresponding to the black-hole mass of the
Schwarzschild--(anti)-de~Sitter solution in the bulk, $H \equiv \dot a/a$ is
the Hubble parameter on the brane, and the term $\kappa / a^2 $ corresponds to
the spatial curvature on the brane. The length scale $\ell$ is defined as
\begin{equation} \label{ell}
\ell = {2 m^2 \over M^3} \, .
\end{equation}

The ``$\pm$'' signs in (\ref{solution}) correspond to two different branches of
the braneworld solutions \cite{SS}.  Models with the lower (``$-$'') sign were
called Brane\,1, and models with the upper (``$+$'') sign were called Brane\,2
in \cite{SS}, and we refer to them in this way throughout this paper.

In what follows, we consider a spatially flat universe ($\kappa = 0$) without
dark radiation ($C = 0$). It is convenient to introduce the dimensionless
cosmological parameters
\begin{equation} \label{omegas}
\Omega_{\rm m} =  {\rho_0 \over 3 m^2 H_0^2} \, , \quad \Omega_\sigma = {\sigma
\over 3 m^2 H_0^2} \, , \quad \Omega_\ell = {1 \over \ell^2 H_0^2} \, , \quad
\Omega_{\Lambda_{\rm b}} = - {\Lambda_{\rm b} \over 6 H_0^2} \, ,
\end{equation}
where the subscript ``{\small 0}'' refers to the current values of cosmological
quantities. The cosmological equation with the energy density $\rho$ dominated
by dust-like matter can now be written in a transparent form:
\begin{equation} \label{hubble0}
\fl {H^2(z) \over H_0^2} = \Omega_{\rm m} (1\!+\!z)^3 + \Omega_\sigma + 2
\Omega_\ell \pm 2 \sqrt{\Omega_\ell}\, \sqrt{\Omega_{\rm m} (1\!+\!z )^3 +
\Omega_\sigma + \Omega_\ell + \Omega_{\Lambda_{\rm b}}} \, .
\end{equation}
The model satisfies the constraint equation
\begin{equation} \label{omega-r1}
\Omega_{\rm m} + \Omega_\sigma \pm 2 \sqrt{\Omega_\ell}\, \sqrt{1 +
\Omega_{\Lambda_{\rm b}}} = 1
\end{equation}
reducing the number of independent $\Omega$ parameters.  The sign choices in
Eqs.~(\ref{hubble0}) and (\ref{omega-r1}) always correspond to each other if $1
+ \Omega_{\Lambda_{\rm b}} > \Omega_\ell\,$.  In the opposite case, $1 +
\Omega_{\Lambda_{\rm b}} < \Omega_\ell\,$, both signs in (\ref{omega-r1})
correspond to the lower sign in (\ref{hubble0}), and the option with both upper
signs in Eqs.~(\ref{hubble0}) and (\ref{omega-r1}) does not exist (see the
appendix of the first paper in \cite{SS} for details).  The signs in
(\ref{omega-r1}) correspond to the two possible ways of bounding the
Schwarzschild--(anti)-de~Sitter bulk space by the brane \cite{CH,Deffayet}.

In the ``normal'' case $1 + \Omega_{\Lambda_{\rm b}} > \Omega_\ell\,$,
substituting $\Omega_\sigma$ from (\ref{omega-r1}) into (\ref{hubble0}), we get
\ber
{H^2(z) \over H_0^2} &=& \Omega_{\rm m} (1\!+\!z)^3 + 1 -  \Omega_{\rm m} +
2 \Omega_\ell \mp 2\sqrt{\Omega_\ell}\, \sqrt{1+\Omega_{\Lambda_{\rm b}}}\nonumber\\
&\pm& 2 \sqrt{\Omega_\ell}\, \sqrt{\Omega_{\rm m} (1\!+\!z)^3 - \Omega_{\rm m}
+ \left( \sqrt{1+\Omega_{\Lambda_{\rm b}}} \mp \sqrt{\Omega_\ell} \right)^2} \,
. \label{hubble1}
\eer

In the ``special'' case $1 + \Omega_{\Lambda_{\rm b}} < \Omega_\ell\,$, the
Brane\,1 model [the lower sign in (\ref{hubble0})] corresponds to both signs of
(\ref{omega-r1}), and the cosmological equation reads as follows:
\ber
{H^2(z) \over H_0^2} &=& \Omega_{\rm m} (1\!+\!z)^3 + 1 -  \Omega_{\rm m} +
2 \Omega_\ell \mp 2\sqrt{\Omega_\ell}\, \sqrt{1+\Omega_{\Lambda_{\rm b}}}\nonumber\\
&-& 2 \sqrt{\Omega_\ell}\, \sqrt{\Omega_{\rm m} (1\!+\!z)^3 - \Omega_{\rm m} +
\left( \sqrt{\Omega_\ell}  \mp \sqrt{1+\Omega_{\Lambda_{\rm b}}} \right)^2} \,
. \label{hubble2}
\eer

For sufficiently high redshifts, the first term on the right-hand side of
(\ref{hubble1}) or (\ref{hubble2}) dominates, and the model reproduces the
matter-dominated Friedmann universe with the density parameter $\Omega_\m$. Now
we note that, for the values of $z$ and parameters $\Omega_{\Lambda_{\rm b}}$
and $\Omega_\ell$ which satisfy
\beq
\Omega_{\rm m} (1\!+\!z)^3 \ll \left( \sqrt{1+\Omega_{\Lambda_{\rm b}}} \mp
\sqrt{\Omega_\ell} \right)^2 \, , \label{eq:mimic0}
\eeq
both Eqs.~(\ref{hubble1}) and (\ref{hubble2}) can be well approximated as
\beq \label{hubble3}
\fl {H^2(z) \over H_0^2} \simeq \Omega_{\rm m} (1\!+\!z)^3 + 1 - \Omega_{\rm m}
- \frac{\sqrt{\Omega_\ell}}{\sqrt{\Omega_\ell} \mp \sqrt{1 +
\Omega_{\Lambda_{\rm b}}}} \left[ \Omega_{\rm m} (1\!+\!z)^3 - \Omega_{\rm m}
\right]\, .
\eeq

We introduce the positive parameter $\alpha$ by the equation
\begin{equation} \label{alpha}
\alpha = {\sqrt{1 + \Omega_{\Lambda_\b}} \over \sqrt{\Omega_\ell}} \, .
\end{equation}
Then, defining a new density parameter by the relation
\beq \label{lcdm}
\Omega^{\rm LCDM}_{\rm m} = {\alpha \over \alpha \mp 1}\, \Omega_{\rm m}  \, ,
\eeq
we get
\beq
{H^2(z) \over H_0^2} \simeq \Omega^{\rm LCDM}_{\rm m} (1\!+\!z)^3 + 1 -
\Omega^{\rm LCDM}_{\rm m}\, ,
\label{eq:lcdm}
\eeq
which is precisely the Hubble parameter describing a LCDM universe. [Note that
the braneworld parameters $\Omega_\ell$ and $\Omega_{\Lambda_{\rm b}}$ have
been effectively absorbed into a ``renormalization'' of the matter density
$\Omega_{\rm m} \to \Omega^{\rm LCDM}_{\rm m}$, defined by (\ref{lcdm}). Since
the value of the parameter $\Omega^{\rm LCDM}_{\rm m}$ should be positive by
virtue of the cosmological considerations, in the case of the upper sign in
(\ref{hubble3}) and (\ref{lcdm}), we restrict ourselves to the parameter region
$\alpha > 1$, i.e., we exclude the special case of Brane\,1 model with the
upper sign in (\ref{omega-r1}) from consideration.]

Thus, our braneworld displays the following remarkable behaviour which we refer
to as {\em ``cosmic mimicry''}:

\begin{itemize}

\item A Brane\,1 model, which at high redshifts expands with density parameter
$\Omega_\m$, at lower redshifts {\em masquerades as a LCDM universe\/} with a
{\em smaller value\/} of the density parameter. In other words, at low
redshifts, the Brane\,1 universe expands as the LCDM model (\ref{eq:lcdm}) with
$\Omega^{\rm LCDM}_{\rm m} < \Omega_{\rm m}$ [where $\Omega^{\rm LCDM}_{\rm m}$
is determined by (\ref{lcdm}) with the lower (``$+$'') sign].

\item A Brane\,2 model at low redshifts also masquerades as LCDM but with a
{\em larger value\/} of the density parameter. In this case, $\Omega^{\rm
LCDM}_{\rm m} > \Omega_{\rm m}$ with $\Omega^{\rm LCDM}_{\rm m}$ being
determined by (\ref{lcdm}) with the upper (``$-$'') sign.

\end{itemize}

The range of redshifts over which this cosmic mimicry occurs is given by $0
\leq z \ll z_{\m}$, with $z_{\m}$ determined by (\ref{eq:mimic0}).
Specifically,
\beq
z_{\m} = \frac{ \left(\sqrt{1 + \Omega_{\Lambda_{\rm b}}} \mp
\sqrt{\Omega_\ell} \right)^{2/3}} {\Omega_{\rm m}^{1/3}} - 1 \, ,
\label{eq:mimic1}
\eeq
which can also be written as
\beq
(1+z_{\m})^3 = {\Omega_\m \left(1 + \Omega_{\Lambda_{\rm b}} \right) \over
\left(\Omega_\m^{\rm LCDM} \right)^2} \label{eq:new1}
\eeq
for both braneworld models.

\begin{figure}
\begin{center}
\includegraphics[width=0.6\textwidth,angle=-90]{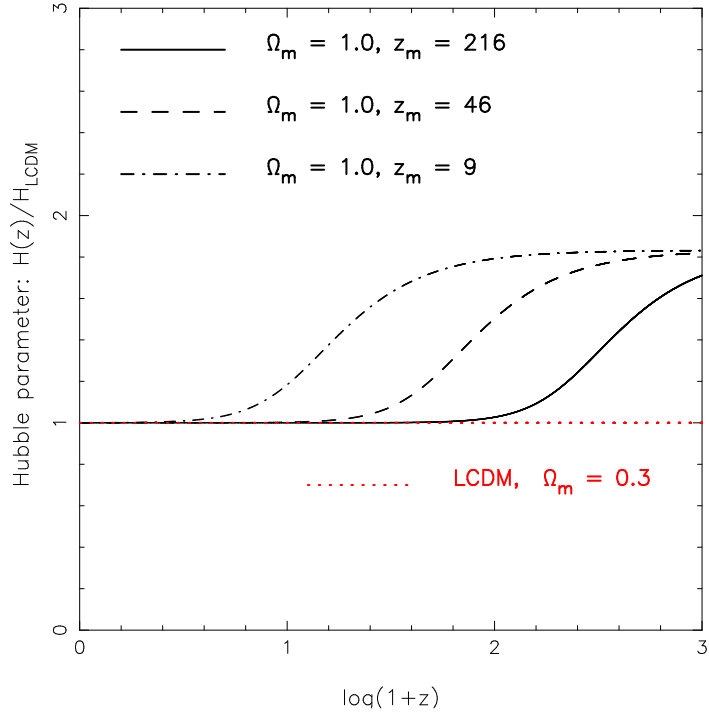}
\end{center}
\caption{\label{fig:fig1}An illustration of cosmic mimicry for the Brane\,1
model. The Hubble parameter in three high-density Brane\,1 models with
$\Omega_\m = 1$ is shown. Also shown is the Hubble parameter in the LCDM model
(dotted line) which closely mimics this braneworld but has a lower mass density
$\Omega_\m^{\rm LCDM} = 0.3$ ($\Omega_\l = 0.7$). The brane matter density
($\Omega_\m$) and the matter density in LCDM are related through $\Omega_\m =
\Omega_\m^{\rm LCDM}\times \left[ 1 + \sqrt{\Omega_\ell/\left(1 + \Omega_{\l_b}
\right)} \right] $, so that $\Omega_\m \gsim \Omega_\m^{\rm LCDM}$. The
redshift interval during which the braneworld masquerades as LCDM, so that
$H_{\rm BRANE1} = H_{\rm LCDM}$ for $z \ll z_\m$, is $z_\m = 9$, $46$, $216$
(left to right) for the three braneworld models.}
\end{figure}

\begin{figure}
\begin{center}
\includegraphics[width=0.6\textwidth,angle=-90]{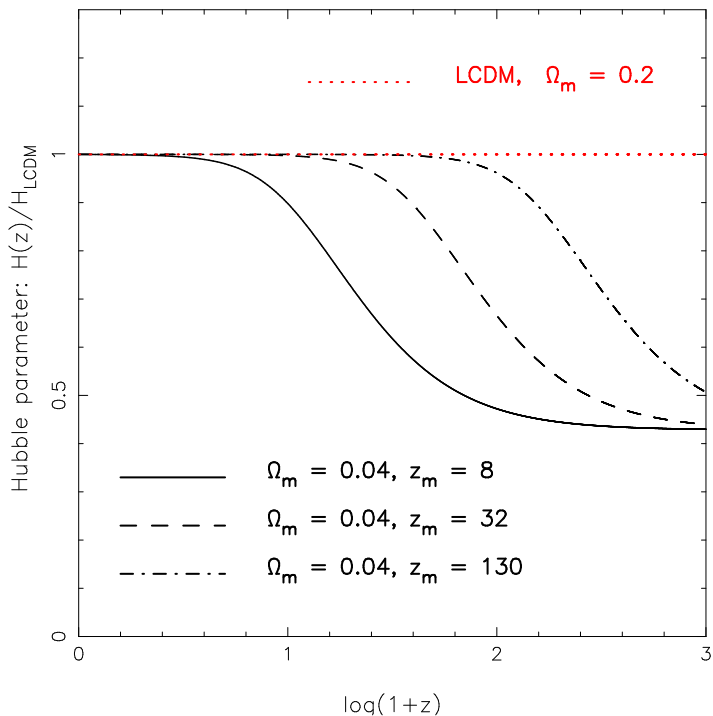}
\end{center}
\caption{\label{fig:fig3}An illustration of cosmic mimicry for the Brane\,2
model. The Hubble parameter in three low-density Brane\,2 models with
$\Omega_\m = 0.04$ is shown. Also shown is the Hubble parameter in a LCDM model
(dotted line) which mimics this braneworld but has a higher mass density
$\Omega_\m^{\rm LCDM} = 0.2$ ($\Omega_\l = 0.8$). The brane matter-density
parameter ($\Omega_\m$) and the corresponding parameter of the masquerading
LCDM model are related as $\Omega_\m = \Omega_\m^{\rm LCDM} \times \left[ 1 -
\sqrt{ \Omega_\ell / \left(1 + \Omega_{\l_b} \right) } \right]$, so that
$\Omega_\m \lsim \Omega_\m^{\rm LCDM}$. The redshift interval during which the
braneworld masquerades as LCDM, so that $H_{\rm BRANE2} = H_{\rm LCDM}$ for $z
\ll z_\m$, is $z_\m = 8$, $32$, $130$ (left to right) for the three braneworld
models.}
\end{figure}

Examples of cosmic mimicry are shown in Fig.~\ref{fig:fig1} for the Brane\,1
model, and in Fig.~\ref{fig:fig3} for the Brane\,2 model. One striking
consequence of Fig.~\ref{fig:fig3} is that a low-density ($\Omega_\m = 0.04$)
universe consisting {\em entirely\/} of baryons mimics a higher-density LCDM
model ($\Omega^{\rm LCDM}_\m = 0.2$) and can therefore be in excellent
agreement with the SN data.

In view of relation (\ref{eq:new1}), it is interesting to note that we can use
the equations derived in this paper to relate the three free parameters in the
braneworld model: $\left\lbrace \Omega_\ell\,, \Omega_{\Lambda_\b},\Omega_\m
\right\rbrace$ to $\left\lbrace \Omega_\m, z_\m, \Omega_\m^{\rm LCDM}
\right\rbrace$. These relations (which turn out to be the same for Brane\,1 and
Brane\,2 models) are:
\ber
\label{eq:relation}
\displaystyle \frac{1 + \Omega_{\Lambda_\b}}{\Omega_\m^{\rm LCDM}} &=&
\displaystyle \frac{\Omega_\m^{\rm
LCDM}}{\Omega_\m} (1+z_\m)^3 \, , \bigskip \\
\displaystyle \frac{\Omega_\ell}{\Omega_\m^{\rm LCDM}} &=& \displaystyle \left[
\sqrt{\frac{\Omega_\m^{\rm LCDM}}{\Omega_\m}} -
\sqrt{\frac{\Omega_\m}{\Omega_\m^{\rm LCDM}}} \right]^2 (1+z_\m)^3 \, .
\eer

Furthermore, if we assume that the value of $\Omega_\m^{\rm LCDM}$ is known
(say, from the analysis of SN data), then the two braneworld parameters
$\Omega_\ell$ and $\Omega_{\Lambda_\b}$ can be related to the two parameters
$\Omega_\m$ and $z_\m$ using (\ref{eq:relation}), so it might be more
convenient to analyze the model in terms of $\Omega_\m$ and $z_\m$ (instead of
$\Omega_\ell$ and $\Omega_{\Lambda_\b}$).

We also note that, under condition (\ref{eq:mimic0}), the brane tension
$\sigma$, determined by (\ref{omega-r1}), is positive for Brane\,1 model, and
negative for Brane\,2 model.

Since the Hubble parameter in braneworld models departs from that in LCDM at
{\em intermediate\/} redshifts ($z > z_{\rm m}$), this could leave behind an
important cosmological signature especially since several key cosmological
observables depend upon the Hubble parameter either differentially or
integrally. Examples include:

\begin{itemize}

\item the luminosity distance $d_L(z)$:
\beq
{d_L(z) \over 1 + z} = c\int_0^z {dz' \over H(z')} \, ,
\label{eq:lumdis}
\eeq

\item the angular-size distance
\beq
d_A(z) = \frac{c}{1+z}\int_0^z {dz' \over H(z')} \, ,
\label{eq:angdis}
\eeq

\item the product $d_A(z)H(z)$, which plays a key role in the Alcock--Paczynski
anisotropy test \cite{alcock},

\item the product $d_A^2(z)H^{-1}(z)$, which is used in the volume-redshift test
\cite{davis},

\item the deceleration parameter:
\beq \label{eq:decel}
q(z) = \frac{H'(z)}{H(z)} (1+z) - 1 \, ,
\eeq

\item the effective equation of state of dark energy:
\beq \label{eq:state0}
w(z) = {2 q(z) - 1 \over 3 \left[ 1 - \Omega_{\rm m}(z) \right] } \, , \quad
\Omega_\m (z) = \Omega_\m \left[ {H_0 \over H (z)} \right]^2 (1 + z)^3 \, ,
\eeq

\item the age of the universe:
\beq\label{eq:age}
t(z) = \int_z^\infty \frac{dz'}{(1+z') H(z')} \, ,
\eeq

\item the ``statefinder pair'' \cite{sssa02}:
\begin{equation}\begin{array}{l}
r = \displaystyle \frac{\stackrel{...}{a}}{a H^3} \equiv  1 + \left\lbrack
\frac{H''}{H} + \left (\frac{H'}{H}\right )^2 \right\rbrack
(1+z)^2 - 2\frac{H'}{H}(1+z) \, , \medskip  \\
s = \displaystyle \frac{r - 1}{3(q - 1/2)} \, , \label{eq:state}
\end{array}
\end{equation}

\item the electron-scattering optical depth to a redshift $z_{\rm reion}$
\beq
\tau(z_{\rm reion}) = c\int_0^{z_{\rm reion}}\frac{n_e(z)\sigma_T
~dz}{(1+z)H(z)} \, , \label{eq:reion}
\eeq
where $n_e$ is the electron density and $\sigma_T$ is the Thomson cross-section
describing scattering between electrons and CMB photons.

\end{itemize}

A degree of caution should be exercised when comparing the late-time LCDM
behaviour (\ref{eq:lcdm}) of our model with different sets of observations,
since the parameter $\Omega^{\rm LCDM}_\m$, residing in (\ref{eq:lcdm}), which
is effectively used in determinations of the luminosity distance
(\ref{eq:lumdis}) and angular-size distance (\ref{eq:angdis}), may very well be
different from the value of $\Omega_\m$ inferred from observations of
gravitational clustering. These issues should be kept in mind when performing a
maximum-likelihood analysis using data belonging to different observational
streams.

\begin{figure}
\begin{center}
\includegraphics[width=0.6\textwidth]{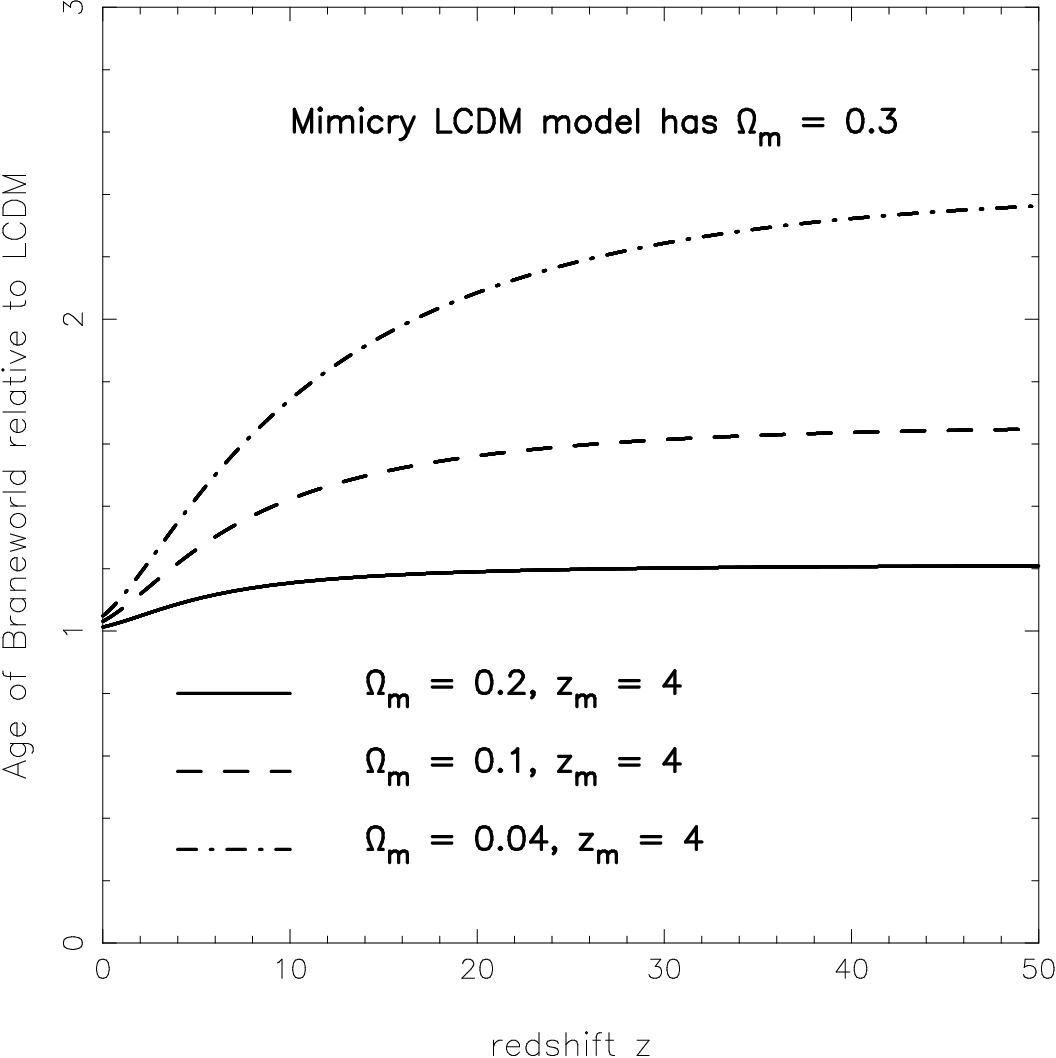}
\end{center}
\caption{\label{fig:age}The age of the universe in the Brane\,2 model is shown
with respect to the LCDM value. The mimicry redshift (\ref{eq:new1}) is $z_\m =
4$ so that $H_{\rm brane}(z) \simeq H_{\rm LCDM}(z)$ at $ z \ll 4$. The
braneworld models have $\Omega_\m = 0.2$, $0.1$, $0.04$ (bottom to top) whereas
$\Omega_\m^{\rm LCDM} = 0.3$. Note that the braneworld models are older than
LCDM.}
\end{figure}

Cosmological tests based on the luminosity distance and angular-size distance
typically probe lower redshifts $z \lsim 2$. Therefore, if the mimicry redshift
is $z_\m \geq 2$, the braneworld model will, for all practical purposes, be
indistinguishable from the LCDM cosmology on the basis of these tests alone.
However, tests which probe higher redshifts should be able to distinguish
between these models. For instance, since $H(z) < H_{\rm LCDM}(z)$ in Brane\,2
at redshifts larger than the mimicry redshift, it follows that the age of the
universe will be greater in this model than in the LCDM cosmology. This is
illustrated in Fig.~\ref{fig:age} for three distinct values of the cosmological
density parameter: $\Omega_\m = 0.2$, $0.1$, $0.04$, all of which are lower
than $\Omega_\m^{\rm LCDM} = 0.3$.

Since the late-time evolution of the universe is
\beq
t(z) \simeq {2 \over 3H_0 \sqrt{\Omega_{\rm m}} } (1+z)^{-3/2} \, ,
\eeq
one finds, for $z \gg 1$,
\beq
\frac{t_{\rm brane}}{t_{\rm LCDM}}(z) \simeq \sqrt{\frac{\Omega_\m^{\rm
LCDM}}{\Omega_\m}} \, .
\eeq
Since $\Omega_\m < \Omega_\m^{\rm LCDM}$ in the Brane\,2 model, we find that
the age of a Brane\,2 universe is greater than that of a LCDM universe. (The
reverse is true for the Brane\,1 model, for which $\Omega_{\rm m} >
\Omega_\m^{\rm LCDM}$.)

The altered rate of expansion in the braneworld model at late times ($z >
z_\m$) also affects other cosmological quantities including the redshift of
reionization which, for the Brane\,2 model, becomes smaller than that in the
LCDM cosmology. This is because the lower value of $H(z)$ in the Brane\,2 model
(relative to the LCDM model), when substituted to (\ref{eq:reion}), gives a
correspondingly lower value for $z_{\rm reion}$ for an identical value of the
optical depth $\tau$ in both models. (In fact, it is easy to see that, for the
Brane\,2 model, the value of $z_{\rm reion}$ decreases with decreasing $z_\m$
and $\Omega_\m$.) Hence, braneworld cosmology makes it possible to have a value
for the reionization redshift which is lower than $z \simeq 17$ inferred for
the LCDM model from the data of the Wilkinson Microwave Anisotropy Probe (WMAP)
\cite{wmap}.

Both an increased age of the universe and a lower redshift of reionization are
attractive properties of the braneworld model which, as we have seen, mimics
the LCDM cosmology at lower redshifts $z < z_\m$.\footnote[3]{Note that the
decreased redshift of reionization and the increased age of the universe are
properties that the Brane\,2 model shares with the loitering universe discussed
in \cite{loiter}.} It is important to note that the presence of high-redshift
quasi-stellar objects (QSO's) and galaxies at redshifts $z \gsim 6$ indicates
that the process of structure formation was already in full swing at that early
epoch when the LCDM universe was less than a billion years old. Most models of
QSO's rely on a central supermassive black hole ($M_{\rm BH} \gsim
10^9M_\odot$) to power the quasar luminosity via accretion. Since structure
forms hierarchically in the cold dark matter scenario, the presence of such
supermassive black holes at high redshift suggest that they formed through an
assembly mechanism involving either accretion or mergers or both. It is not
clear whether either of these processes is efficient enough to assemble a large
number of high-redshift QSO's in a LCDM cosmology \cite{richards03,bh}. (At the
time of writing, over 400 quasi-stellar objects with redshifts $\geq 4$ have
been reported in the literature.)  In addition, it is interesting to note the
discovery of an ``old'' quasar APM~08279+5255 at redshift $z = 3.91$, which
appears to have an older age (2.1~Gyr) than the LCDM model (at that redshift)
and may, therefore, be problematic for concordance cosmology as well as
several other dark-energy models \cite{quasar}.  Since Brane\,2 can be
considerably older than LCDM, this makes it easier to account for the existence
of APM~08279+5255 within the braneworld context than in LCDM.  (We shall return
to this issue in a companion paper.)

Another ``surprise'' for the concordance cosmology emerges from the WMAP data
which suggest that the optical depth to reionization is $\tau = 0.17 \pm 0.06$.
For the LCDM model, this translates into a rather high redshift of reionization
$z = 17 \pm 5$ \cite{wmap}, suggesting that population~III stars and/or
mini-QSO's were already in place at $z > 17$ in order to have efficiently
reionized the universe. Whether both the structure-assembly issue and the high
reionization redshift can be successfully accommodated within the LCDM paradigm
remains to be seen \cite{reionization}. However, both issues are important
enough to compel the serious theorist to look for alternative models which,
while preserving the manifold strengths and successes of the LCDM model at low
$z$, will also be able to flexibly accommodate the striking properties of the
universe at higher redshifts. As we have demonstrated in this paper, braneworld
cosmology may successfully alleviate some of the tension currently existing
between theory and observations at moderate redshifts, while allowing the
universe to be ``LCDM-like'' at the present epoch.

\begin{figure}
\begin{center}
\includegraphics[width=0.6\textwidth]{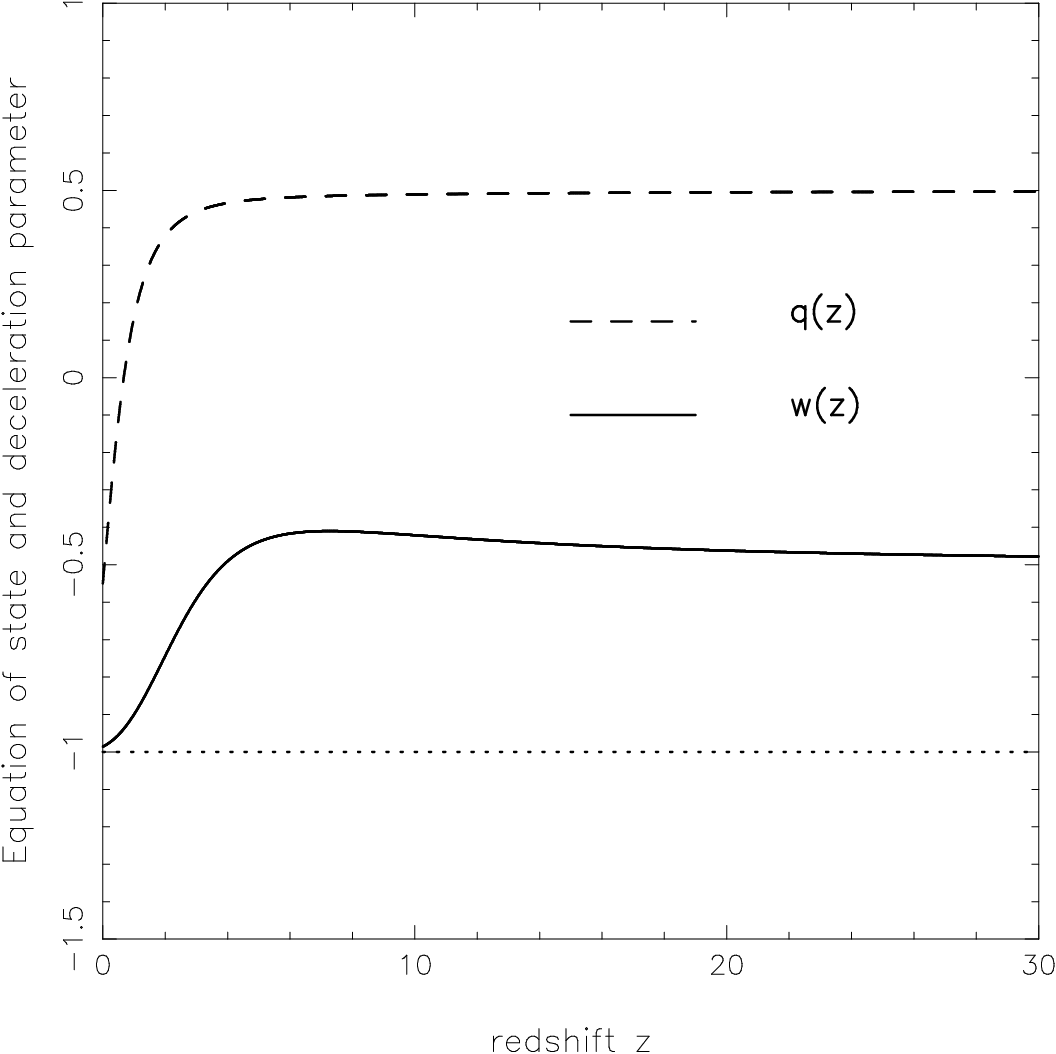}
\end{center}
\caption{\label{fig:state}The effective equation of state (solid line) and the
deceleration parameter (dashed line) of the Brane\,2 model are shown. (The
dotted line shows $w=-1$ which describes the LCDM model.) The mimicry redshift
(\ref{eq:new1}) is $z_\m = 4$ so that $H_{\rm brane}(z) \simeq H_{\rm LCDM}(z)$
at $ z \lsim 4$. The braneworld has $\Omega_\m = 0.2$ whereas $\Omega_\m^{\rm
LCDM} = 0.3$.}
\end{figure}

The effective equation of state and the deceleration parameter of the Brane\,2
model are shown in Fig.~\ref{fig:state}. The braneworld has $\Omega_\m = 0.2$
and, at $z \lsim 4$, masquerades as a higher-density LCDM model with
$\Omega_\m^{\rm LCDM} = 0.3$.  Note that the {\em effective\/} equation of
state (\ref{eq:state0}) is a {\em model-dependent\/} quantity, involving the
model-dependent cosmological parameter $\Omega_\m$ in its definition.  In our
case, we use the braneworld theory as our model with $\Omega_\m$ defined in
(\ref{omegas}), and the effective equation of state (\ref{eq:state0}) is then
{\em redshift-dependent\/} even during the mimicry period when $H_{\rm brane}
\bigl( \Omega_\m, z \bigr) \simeq H_{\rm LCDM} \bigl( \Omega_\m^{\rm LCDM},z
\bigr)$. A theorist who is unaware of the possibility of cosmic mimicry, when
reconstructing the cosmic equation of state from (\ref{eq:state0}) with
$\Omega_\m^{\rm LCDM} = 0.3$ in the place of $\Omega_\m$, will arrive at a
different conclusion that $w = -1$. This example demonstrates some of the
pitfalls associated with the cosmological reconstruction of the equation of
state, which depends on the underlying theoretical model and for which an
accurate knowledge of $\Omega_\m$ is essential; see
\cite{sssa02,maor02,alam04,bassett04} for a discussion of related issues.

In concluding this section, we would like to mention an additional important
consequence of braneworld cosmology, namely that the departure of the brane
Hubble parameter from its LCDM counterpart is likely to affect the growth of
density perturbations. From the linearized perturbation equation
\beq
{\ddot \delta} + 2H{\dot\delta} - 4\pi G{\bar\rho}~\delta = 0 \, ,
\label{eq:delta}
\eeq
where $\delta = (\rho-{\bar\rho})/{\bar\rho}$ is the relative density
perturbation, it follows that a large value of $H$ will, via the second term on
the left-hand side, dampen the growth of perturbations, while a small value of
$H$ should hasten their growth. Although this argument is also likely to carry
over to the braneworld, the direct application of (\ref{eq:delta}) to the study
of gravitational instability (as well as CMB anisotropy) should be treated with
some caution since (\ref{eq:delta}) was derived within the general-relativistic
framework, and it is likely to require some modification before it can be
applied to the study of cosmological perturbations on the brane. Indeed,
according to the analysis of \cite{ls,lss}, the linearized equation describing
the evolution of inhomogeneities in the braneworld models of
Dvali--Gabadadze--Porrati (DGP) type \cite{DGP}, corresponding to $\Lambda_{\rm
b} = 0$, on spatial scales much shorter than $\ell$ has the form
\begin{equation} \label{perturb}
{\ddot \delta} + 2H{\dot\delta} - 4\pi G{\bar\rho}\left(1 +
\frac{1}{3\beta_*}\right) \delta = 0 \, , \label{eq:delta1}
\end{equation}
where the time-dependent parameter $\beta_*$ is given by \cite{ls}
\begin{equation} \label{betastar}
\beta_* (t) = 1 \mp \ell H \left(1 + {\dot H \over 3 H^2} \right) \, .
\end{equation}

Another equation which might possibly describe the behaviour of density
perturbations in the braneworld models under consideration in this paper will
be discussed in the next section.

\section{Properties of braneworld gravity}\label{gravity}

\subsection{Multiscale gravity}

Here, we would like to discuss the results of the previous section in the light
of some generic properties of braneworld gravity.

Cosmology determined by action (\ref{action}) has two important scales, namely,
the length scale (\ref{ell})
\begin{equation}
\ell = {2 m^2 \over M^3} \, ,
\end{equation}
which describes the interplay between the bulk and brane gravity,
and the momentum scale
\begin{equation} \label{k}
k = {\sigma \over 3 M^3} \, ,
\end{equation}
which determines the role of the brane tension in the dynamics of the brane. In
a model characterized by the Randall--Sundrum constraint \cite{RS}
\begin{equation} \label{lambda-rs}
\Lambda_{\rm RS} \equiv {\Lambda_\b \over 2} + {\sigma^2 \over 3 M^6} = 0 \, ,
\end{equation}
the absolute value of $k$ is equal to the inverse warping length $\ell_{\rm
warp} = \sqrt{- 6 / \Lambda_\b}$ of the Randall--Sundrum solution. Note that
$k$ is negative when $\sigma < 0$. We take $M$ to be positive in our
paper, since, in the opposite case, the massive Kaluza--Klein gravitons become
ghosts \cite{Padilla,SVi}.

Following the procedure first employed in \cite{SMS} for the Randall--Sundrum
(RS) model \cite{RS} and subsequently applied in \cite{MMT} to the model under
consideration, we contract once the Gauss identity
\begin{equation}\label{Gauss}
R_{abc}{}^d = h_a{}^f h_b{}^g h_c{}^k h^d{}_j {\cal R}_{fgk}{}^j + K_{ac}
K_b{}^d - K_{bc} K_a{}^d
\end{equation}
on the brane and, using Eq.~(\ref{bulk}), obtain the equation
\begin{equation}\label{effective}
G_{ab} + \Lambda_{\rm eff} h_{ab} = 8 \pi G_{\rm eff} \tau_{ab} + {1 \over
\beta + 1} \left( {1 \over M^6} Q_{ab} - W_{ab} \right) \, ,
\end{equation}
where
\begin{equation}\label{beta}
\beta = {2 \sigma m^2 \over 3 M^6} = k \ell
\end{equation}
is an important dimensionless parameter,
\begin{equation} \label{lambda-eff}
\Lambda_{\rm eff} = {\Lambda_{\rm RS} \over \beta + 1}
\end{equation}
is the effective cosmological constant on the brane,
\begin{equation}\label{g-eff}
8 \pi G_{\rm eff} = {\beta \over \beta + 1} \cdot {1 \over m^2}
\end{equation}
is the effective gravitational constant,
\begin{equation}\label{q}
Q_{ab} = \frac13 E E_{ab} - E_{ac} E^{c}{}_b + \frac12 \left(E_{cd} E^{cd} -
\frac13 E^2 \right) h_{ab}
\end{equation}
is a quadratic expression constructed from the ``bare'' Einstein equation
$E_{ab} \equiv m^2 G_{ab} - \tau_{ab}$ on the brane, and $E = h^{ab} E_{ab}$.
The symmetric traceless tensor $W_{ab} \equiv n^c n^d W_{acbd}$ in
(\ref{effective}) is a projection of the bulk Weyl tensor $W_{abcd}$.  It is
related to the tensor $Q_{ab}$ through the conservation equation
\begin{equation}\label{conserv}
D^a \left( Q_{ab} - M^6 W_{ab} \right) = 0 \, ,
\end{equation}
where $D^a$ denotes the covariant derivative on the brane associated with the
induced metric $h_{ab}$.

It is important to note that all couplings in Eq.~(\ref{effective}), including
the effective cosmological and gravitational constants, are inversely
proportional to $\beta + 1$, which indicates that the theory becomes singular
for the special case $\beta = - 1$ (see \cite{SVi,Smolyakov}).

In the absence of the curvature term on the brane ($m = 0$), we obtain
Eq.~(\ref{effective}) in which $8 \pi G_{\rm eff} = 2 \sigma / 3 M^6$ is the
gravitational constant in the RS model \cite{RS}, and $\beta = 0$; in this
form, Eq.~(\ref{effective}) was first derived in \cite{SMS}. The conditions
$\sigma = 0$ and $\Lambda_\b = 0$ are characteristic of the DGP model
\cite{DGP}, which also has $\beta = 0$. In this model, the effective
gravitational constant (\ref{g-eff}) turns to zero, i.e., the term linear in
the stress--energy tensor on the brane vanishes in Eq.~(\ref{effective}).

Let us now turn to the ``cosmic mimicry'' exhibited by braneworld cosmology and
show how it can be related to the gravitational properties of braneworld
theory. In this context, it is remarkable that the parameter $\beta$ introduced
in (\ref{beta}) is very close (in absolute terms) to the parameter $\alpha$
introduced in (\ref{alpha}) in our discussion of mimicry models. Specifically,
\begin{equation}\label{albeta}
\beta = {1 - \Omega_\m \over 2 \Omega_\ell} \mp \alpha \, ,
\end{equation}
so that
\beq \label{alphabeta}
|\beta| \approx \alpha
\eeq
when $|1 - \Omega_\m| \ll \Omega_\ell\,$. This last inequality follows
naturally from condition (\ref{eq:mimic0}) for values of $\alpha$ of order
unity, which are of interest to us in this paper. As a consequence, the term
which appears in the ``renormalization'' of the cosmological mass density
$\Omega_\m$ in (\ref{lcdm}) is almost identical to the term which redefines the
gravitational constant in (\ref{g-eff}). This coincidence can be explained by
inspecting the brane equation (\ref{effective}). First, we note that
cosmological solutions without dark radiation are embeddable in the
anti-de~Sitter bulk spacetime, so that $W_{ab} = 0$ for these solutions. For
large cosmological matter densities, the quadratic expression (\ref{q})
dominates in Eq.~(\ref{effective}), and the universe is described by the
``bare'' Einstein equation $m^2 G_{ab} - \tau_{ab} = 0$, with the effective
gravitational coupling being equal to $1 / m^2$. As the matter density
decreases, the role of this quadratic term becomes less and less important, and
the effective gravitational coupling eventually is determined by the linear
part of Eq.~(\ref{effective}), i.e., by the gravitational constant
(\ref{g-eff}). Thus, comparing (\ref{g-eff}) and (\ref{lcdm}), one has the
following natural relation, valid to a high precision in view of
(\ref{alphabeta}):
\begin{equation} \label{omega-m}
\Omega_\m^{\rm LCDM} = {8 \pi G_{\rm eff} \rho_0 \over 3 H_0^2 } \, .
\end{equation}

Now we consider some relevant properties of braneworld gravity. The expression
for $Q_{ab}$ in Eq.~(\ref{q}) is quadratic in the curvature as well as in the
stress--energy tensor. On the other hand, the tensor $W_{ab}$ is related to
$Q_{ab}$ through the conservation equation (\ref{conserv}). One might,
therefore, expect that the term in the parentheses on the right-hand side of
Eq.~(\ref{effective}), namely $Q_{ab}/M^6 - W_{ab}$, will be insignificant on
sufficiently large length scales, and that the braneworld theory on those
scales should reduce to Einstein gravity with the effective constants given by
(\ref{lambda-eff}) and (\ref{g-eff}). This expectation is borne out by a
detailed analysis \cite{SVi} carried out for a positive-tension brane ($\sigma
> 0$) in the specific case when the braneworld satisfies the RS constraint
(\ref{lambda-rs}). In this case, the gravitational potential of a unit mass on
large scales (on the positive-tension brane) has the Newtonian form with a
small RS correction \cite{SVi}\footnote{Here, we present the results for the
two-brane model as the second (negative-tension) brane is taken to spatial
infinity.}:
\begin{equation} \label{large-scale}
V(r) = - {G_{\rm eff} \over r} \left[ 1 + {2 \over 3 (\beta + 1) (kr)^2}
\right] \, , \quad kr \gg 1 \, ,
\end{equation}
where $G_{\rm eff}$ is given by (\ref{g-eff}).

On smaller spatial scales, $kr \ll 1, \beta$, the potential in linear theory
again has the Newtonian form, this time with a small logarithmic correction:
\begin{equation} \label{small-scale}
V (r) = - {\widetilde G_{\rm eff} \over r} - \left({15 \over 8} +
\frac{2}{\beta} \right) {k \over 3 \pi^2 m^2} \log \left[\left({15 \over 8} +
\frac2\beta \right) kr \right] \, , \quad kr \ll 1, \beta \, ,
\end{equation}
and with a different expression for the effective gravitational constant
\cite{SVi}
\begin{equation} \label{g-tilde}
\widetilde G_{\rm eff} = \left[ 1 + {1 \over 3 ( 1 + \beta )} \right] {1 \over
8\pi m^2} = \left(1 + {4 \over 3 \beta} \right)  G_{\rm eff} \, .
\end{equation}
For $k \to 0$ (hence, also $\beta \to 0$), this reproduces the result obtained
for the DGP model in \cite{vdvz} on scales $r \ll \ell$.

It is worth noting that gravity on these smaller scales $kr \ll 1, \beta$, in
general, involves the massless scalar radion, i.e., it is of scalar--tensor
type. As a consequence, for the spherically symmetric solution, it violates the
property $h_{00} (r) = {} - h_{rr}^{-1} (r) $, or, in the linear approximation,
$\gamma_{00} (r) = \gamma_{rr} (r)$, where $\gamma_{\alpha\beta} (r)$ are the
components of metric perturbation in the spherically symmetric coordinate
system. Specifically, in the model with the RS constraint one can obtain the
relation:
\begin{equation}\label{discrep}
{\Delta \gamma \over \gamma_{00}} \equiv {\gamma_{00} - \gamma_{rr} \over
\gamma_{00}} = {1 \over 1 + 3 \beta / 4} \, .
\end{equation}
Since there are stringent experimental upper bounds \cite{gamma} on the
left-hand side of (\ref{discrep}) in the neighbourhood of the solar system (it
should not exceed $10^{-5}$ by order of magnitude), if solution
(\ref{small-scale}) were applicable in this domain, it would imply that only
very large values of $\beta$ are permissible in the braneworld theory under
consideration [namely, the braneworld model (\ref{solution}) with the RS
constraint (\ref{lambda-rs})]. It is important to note, however, that the
applicability region of the linear approximation (\ref{small-scale}) is bounded
from below by a length scale which depends upon the mass of the central source,
as has been demonstrated for the DGP model in \cite{vdvz}. Specifically, the
dynamics of the radion develops strong nonlinear corrections on sufficiently
small scales, leading to the breakdown of linearized theory. (This also creates
the so-called strong-coupling problem in the DGP model \cite{problems}.) In
this case, in order to study gravity at small distances from the source, one
must turn to the fully nonlinear theory.

Next, let us determine the distances at which the linearized theory, leading to
(\ref{small-scale}), breaks down and to establish the correct behaviour of the
potential on such scales.  In order to do this, we turn to the effective
equation (\ref{effective}). Taking the trace of Eq.~(\ref{effective}), we get
the following closed scalar equation on the brane:
\begin{equation}\label{trace}
{} - R + 4 \Lambda_{\rm eff} - 8 \pi G_{\rm eff} \tau = {Q \over (\beta + 1)
M^6} \, ,
\end{equation}
where the left-hand side contains terms which are linear in the curvature and
in the stress--energy tensor while the right-hand side contains the quadratic
term $Q = h^{ab} Q_{ab}$.

Suppose that we are interested in the behaviour of gravity in the neighbourhood
of a spherically symmetric source with density $\rho_s$, total mass ${\cal
M}_s$, and radius $r_s$.  First of all, we assume that one can neglect the
tensor $W_{ab}$ and the effective cosmological constant in the neighbourhood of
the source. As regards the effective cosmological constant, this assumption is
natural. Concerning the tensor $W_{ab}$, its smallness in the neighbourhood of
the source represents some additional condition on the spherically symmetric
solution. We believe that a condition of this sort is likely to arise in any
consistent and viable braneworld theory since, without it, one has a large number
of spherically symmetric solutions on the brane, many of them non-physical (see
\cite{VW} for a comprehensive treatment in the framework of the RS model).

Within the source itself, we have two qualitatively different options: an
approximate solution can be sought either neglecting the quadratic part or
linear part of Eqs.~(\ref{effective}) and (\ref{trace}).  We should choose the
option that gives the smaller error of approximation in Eq.~(\ref{trace}). In the
first case, neglecting the quadratic part and the effective cosmological
constant, we have
\begin{equation}\label{lin}
G_{ab} - 8 \pi G_{\rm eff} \tau_{ab} \approx 0 \quad \Rightarrow \quad {Q \over
(\beta + 1) M^6} \sim {\rho_s^2 \over (\beta + 1)^3 M^6} \, .
\end{equation}
In the second case, we neglect the linear part, so that
\begin{equation}\label{quadrat}
Q_{ab} \approx 0 \quad \Rightarrow \quad E_{ab} \approx 0 \quad \Rightarrow
\quad  R + 8 \pi G_{\rm eff} \tau \sim {\rho_s \over (\beta + 1) m^2 } \, .
\end{equation}
The final expression on the right-hand side of (\ref{quadrat}) is smaller than
the corresponding expression in (\ref{lin}) if
\begin{equation}\label{scale}
\rho_s > (\beta + 1)^2 {M^6 \over m^2} \quad \Rightarrow \quad r_s^3 < r_*^3
\sim {{\cal M}_s \ell^2 \over (\beta + 1)^2 m^2 } \, ,
\end{equation}
where we used the relation ${\cal M}_s \sim \rho_s r_s^3$. Thus, we can expect
that, in the neighbourhood of the source, on distances smaller than $r_*$ given
by (\ref{scale}), the solution is determined mainly by the quadratic part
$Q_{ab}$ in Eq.~(\ref{effective}), which means that it respects the ``bare''
Einstein equation $m^2 G_{ab} = \tau_{ab}$ to a high precision.  This effect is
sometimes described as the ``gravity filter'' of the DGP model \cite{DGP},
which screens the scalar graviton in the neighbourhood of the source making the
gravity effectively Einsteinian. Some aspects of this interesting phenomenon
are discussed in recent papers \cite{Kaloper}.

Expression (\ref{scale}) generalizes the length scale \cite{vdvz} of the DGP
model, below which nonlinear effects become important, to the case of nonzero
brane tension (nonzero $\beta$) and bulk cosmological constant satisfying the
RS constraint (\ref{lambda-rs}). The observable gravitational constant on
scales much smaller than $r_*$ will be given by
\begin{equation}\label{g-obs}
8 \pi G_{\rm obs} = {1 \over m^2} \, .
\end{equation}
For the Sun, the scale $r_*$ is estimated as
\begin{equation}
r_* \equiv r_{\odot} \sim {10^{16}\, \mbox{km} \over (\beta + 1)^{2/3}\,
\Omega_\ell^{1/3} } \, ,
\end{equation}
which, for interesting values of $\beta$ and $\Omega_\ell\,$, will be very
large. The corresponding radius for the Earth is smaller only by two orders of
magnitude.

This, however, is not the full story.  As argued in \cite{ls,lss}, the
gravitational potential of a spherically symmetric body on scales $r_* \lsim r
\ll \ell$ is corrected by the cosmological expansion. Moreover, the critical
scale $r_*$ becomes dependent on the value of the Hubble parameter, and can be
different from (\ref{scale}) for values of $\ell$ of the order of the Hubble
length. Another effective gravitational constant appears in the equation for
the growth of cosmological perturbations on the above-mentioned scales, as was
mentioned in the previous section [see Eqs.~(\ref{perturb}) and
(\ref{betastar})]. However, we note that, in this paper, interesting values of
the parameter $\ell$ are smaller than the Hubble length since we would like
relation (\ref{eq:mimic0}) to be satisfied for reasonably high redshifts, and
this condition implies that the parameter $\Omega_\ell\,$, defined in
Eq.~(\ref{omegas}), is significantly larger than unity. Consequently, the
critical scale $r_*$ is given by Eq.~(\ref{scale}). Note that gravity on scales
$r \ll r_*\,$, although close to Einstein gravity, is not exactly Einsteinian,
and these deviations can be used to test braneworld theory on solar-system
scales \cite{ls,vdvz,lss,Iorio}.

The appearance of the distance scale $r_*$ given by (\ref{scale}) can also be
justified by using an argument from cosmology.  Imagine the central body to be
formed of pressureless matter and to represent a part of the homogeneous and
isotropic universe. Then the solution inside this body is uniquely described by
the cosmological equations of Sec.~\ref{mimicry}\@.  As the density parameter
of the body exceeds the right-hand side of (\ref{eq:mimic0}), its gravitational
evolution is effectively governed by the ``bare'' Einstein equation $m^2 G_{ab}
= \tau_{ab}$.  The relevant condition describing this case is the inequality
opposite to (\ref{eq:mimic0}); it can be written in terms of the density
$\rho_s$ as
\begin{equation}
\rho_s > (\alpha \mp 1)^2 {M^6 \over m^2} \, ,
\label{inequality}
\end{equation}
which, in view of relations (\ref{albeta}) and (\ref{alphabeta}), essentially
coincides with condition (\ref{scale}).

Although the preceding reasoning is applicable to both the positive-tension and
the negative-tension brane, the current understanding of the braneworld
gravitational physics supports only the positive-tension case (which
corresponds to Brane\,1 model in this paper), while the situation with the
negative-tension brane remains unclear (at least, to the authors of this
paper). From Eq.~(\ref{effective}), one might expect that a negative-tension
brane will show reasonable physical behaviour in the case $|\beta| > 1$ (note
that $\beta < 0$ for a negative-tension brane), in which the gravitational
constant (\ref{g-eff}) is positive. However, direct calculation (along the
lines of \cite{SVi}) {\em in the two-brane model\/} with the RS constraint
(\ref{lambda-rs}) shows that, in this case, the gravitational interaction
between material bodies on large scales is dominated by the ghost-like radion,
with the effective gravitational coupling
\begin{equation}\label{g-radion}
G_{\rm radion} = {} - \frac13\, G_{\rm eff} \, , \quad kr \gg 1 \, ,
\end{equation}
where $G_{\rm eff}$ is given by the same expression (\ref{g-eff}).  The
radion-dominated gravity on these scales is formally attractive in the case
$G_{\rm eff} < 0$, and is repulsive for $G_{\rm eff} > 0$.  However, on smaller
spatial scales $kr_* \lsim kr \ll 1, |\beta|$, Newton's law similar to
(\ref{small-scale}) is reproduced with the gravitational constant given by
(\ref{g-tilde}), which is positive if $|\beta| > 4/3$.  The gravity on these
scales is of scalar--tensor character. On still smaller distances from the
central source, $r < r_*$, the theory may approach Einstein gravity with the
effective gravitational constant (\ref{g-obs}). However, all these expectations
remain to be verified by more refined calculations which generalize to the
braneworld models of interest to us in this paper, including those with
negative tension and without the RS constraint (\ref{lambda-rs}).  In the
following subsection, we propose a different approach to this problem, based on
alternative boundary conditions in the brane--bulk system.

\subsection{The role of boundary conditions in braneworld models}

Many papers (see, e.g., the recent review \cite{Lue}) have stressed that
equation (\ref{effective}) is not closed on the brane in the sense that it
contains the symmetric traceless tensor $W_{ab}$ whose behaviour on the brane
is not determined by the dynamics of matter alone. As of now, this
``nonclosure'' of the equations on the brane constitutes one of the major
problems confronting braneworld theory.  Some additional information from the
bulk is needed to determine the observable braneworld dynamics completely; one
can say that some {\em boundary conditions\/} are to be specified to restrict
the class of possible braneworld solutions.  Usually, one tends to specify the
``physical'' boundary conditions somewhere in the bulk, by requiring the bulk
to be singularity-free (see \cite{Lue}) and/or by introducing one or more
(regulatory) branes in the bulk. However, it seems to be difficult to implement
this proposal in practice in a fully nonlinear theory, and it thus remains
unclear whether such proposals will lead to a reasonable theory at all. A more
practical approach to this problem, adopted in many papers (see, e.g.,
\cite{Maartens}), consists in making some reasonable assumptions about the
tensor $W_{ab}$ on the brane. Here, we make a proposal to take this approach as
a way of specifying the boundary conditions in the brane-bulk system. The
proposal consists in specifying the boundary conditions on the brane, rather
than in the bulk, in such a way as to close equation (\ref{effective}).

Of all possible versions of boundary conditions of this kind, the simplest one
appears to consist in setting
\begin{equation} \label{bc}
W_{ab} = 0
\end{equation}
on the brane.  This condition is very often imposed in practical calculations
by neglecting the contribution of this term in various situations. Here, we do
not discuss the physical meaning of this proposal in all detail but only
consider its consequences very briefly.

With the specification of the boundary condition in the form (\ref{bc}),
equation (\ref{effective}) becomes a closed four-dimensional equation on the
brane completely determining its dynamics, and solution in the bulk becomes of
no crucial importance for an observer on the brane.  Note that the
specification (\ref{bc}) preserves all homogeneous and isotropic cosmological
solutions with the only restriction that the dark radiation is zero, i.e., $C =
0$ in Eq.~(\ref{solution}).  Thus we are not losing the cosmological properties
of the braneworld models (except those for which the nonzero value of the dark
radiation is crucial \cite{loiter}); in particular, the cosmic mimicry
phenomenon discussed in this paper remains intact.  The boundary condition
under consideration affects only the inhomogeneous gravitational physics.

It was shown in \cite{KPP} that a spherically symmetric metric in the case
(\ref{bc}) is unique and is the Schwarzschild metric in a broad region of
parameters of the theory. Thus, in particular, there are no higher-dimensional
corrections to the Newtonian potential, as is the case in the two-brane model
[see Eqs.~(\ref{large-scale}) and (\ref{small-scale})], and the theory is not
``spoiled'' by the presence of the scalar gravitational degree of freedom. The
small-scale gravitational physics in this case has the effective gravitational
constant equal to $1/m^2$, as in the previous subsection. (From this viewpoint,
the two-brane model discussed in the preceding subsection represents a
braneworld model with a different type of boundary condition.)

The linear equations for the cosmological scalar density perturbations in the
braneworld model under consideration with the boundary condition (\ref{bc})
were, in fact, recently derived in \cite{SC}.  Condition (\ref{bc}) in
\cite{SC} was assumed as a reasonable approximation on sufficiently small
spatial scales while, in this paper, we propose to take it as an exact boundary
condition. The equation governing the evolution of the relativistic potential
$\Phi$ was shown in \cite{SC} to have the form
\begin{equation}  \label{scalar}
{1 \over a^2} \nabla^2 \Phi - 3 H^2 \Phi - 3 H \dot \Phi =  - 8 \pi G^*_{\rm
eff} \delta \rho \, ,
\end{equation}
where
\begin{equation}
 8 \pi G^*_{\rm eff} = {1 \over m^2} \left( 1 \pm {1 \over \sqrt{1 + \ell^2
\left( {\rho + \sigma \over 3 m^2} - {\Lambda_\b \over 6} \right)}} \right) = -
{2 \dot H \over \rho} \, ,
\end{equation}
and the last equality is valid for the case of dust.  As in the discussion in
the end of the previous section, again the only change is the appearance of a
time-dependent effective gravitational constant.  Equation (\ref{scalar}) was
used to test the theory against CMB observations in \cite{SC}.

We note that the proposal of the boundary conditions (\ref{bc}) works here for
branes with positive as well as negative tensions and, hopefully, represents
one of the ways of combining the attractive features of the mimicry cosmology
of the Brane\,2 model with the observable small-scale gravity. More analysis is
required to test this proposal, and we are going to return to this issue in the
future publications.

\section{Conclusion}\label{conclude}

We have shown that braneworld cosmology, for a large region of parameter space,
exhibits a property which can be called ``cosmic mimicry.'' During early
cosmological epochs, the braneworld behaves like a matter-dominated Friedmann
universe with the {\em usual\/} value of the cosmological parameter
$\Omega_{\rm m}$ that would be inferred from observations of the local matter
density.  At late times, however, the universe evolves almost exactly like in
the LCDM scenario but with a {\em renormalized\/} value of the cosmological
density parameter $\Omega^{\rm LCDM}_{\rm m}$.  Specifically, a
positive-tension Brane\,1 model, which at high redshifts expands with density
parameter $\Omega_\m$, at lower redshifts mimics the LCDM cosmology with a {\em
smaller value\/} of the density parameter $\Omega^{\rm LCDM}_{\rm m} <
\Omega_{\rm m}$. A negative-tension Brane\,2 model at low redshifts also mimics
LCDM but with a {\em larger value\/} of the density parameter $\Omega^{\rm
LCDM}_{\rm m} > \Omega_{\rm m}$.\footnote{The value of $\Omega^{\rm LCDM}_{\rm
m}$ is determined by (\ref{lcdm}) with the upper (``$-$'') sign for Brane\,2
and the lower (``$+$'') sign for Brane\,1.} The transition redshift between the
early epoch and the late (mimicry) epoch is, in principle, a free parameter
which depends upon constants entering the braneworld Lagrangian; see
(\ref{eq:mimic1}) or (\ref{eq:new1}).

The braneworld models discussed in this paper have interesting cosmological
properties. For instance, in the case of Brane\,1 (Brane\,2), the universe
expands faster (slower) than in the LCDM scenario at redshifts greater than the
mimicry redshift $z_\m$, whereas, for $z < z_\m$, $H_{\rm brane}(z) \equiv
H_{\rm LCDM}(z)$ in both models. The smaller value of the Hubble parameter at
intermediate redshifts ($z > \mbox{few}$) in the case of Brane\,2 leads to an
older universe and also to a redshift of reionization which can be
significantly lower than $z \simeq 17$ inferred for the LCDM model from the
WMAP data \cite{wmap}. These features, together with the mimicry property,
result in an interesting new cosmology which can successfully ameliorate the
current tension between the concordance (LCDM) cosmology and the high-redshift
universe.

The effect of cosmic mimicry and the existence of two asymptotic density
parameters $\Omega_\m$ and $\Omega^{\rm LCDM}_{\rm m}$ is a consequence of the
time-dependence of the effective gravitational constant in braneworld theory
\cite{MMT}, which can be related to the well known property of the
scale-dependence of the effective gravitational constant in braneworld models
\cite{vdvz}. On large spatial scales, $kr \gg 1$, the braneworld model with
positive brane tension (Brane\,1) exhibits gravity with the renormalized
effective gravitational constant (\ref{g-eff}), and we showed that this
renormalization corresponds to the renormalization of the cosmological density
parameter (\ref{omega-m}).  The behaviour of the braneworld model with negative
brane tension (Brane\,2) on these scales is drastically different: it is
dominated by the scalar radion which can even make the gravitational
interaction repulsive; see Eq.~(\ref{g-radion}).\footnote{This result was
obtained for the brane with the RS constraint, but the presence of a small
effective cosmological constant (\ref{lambda-eff}) is unlikely to remedy the
situation.} The problem of this kind may not arise in models with stabilized
radion \cite{GW} or in models with different boundary conditions for the
brane--bulk system --- but this issue has to be studied separately.

Only two effective gravitational constants appear in the cosmology under
consideration, given by (\ref{g-eff}) and (\ref{g-obs}), respectively, for low
and high energy densities. However, in the local gravitational physics, there
also appears the spatial distance (\ref{scale}) depending upon the mass of the
central source, so that gravity in the range
\begin{equation}
r_* \lsim r \lsim \ell
\end{equation}
of distances $r$ from the source has a different value of the gravitational
constant, given by (\ref{g-tilde}), and, moreover, typically has a
scalar--tensor character manifested, in particular, in (\ref{discrep}).  This
may be important for the estimates of masses from the dynamics of clusters of
galaxies and from gravitational lensing on these scales in the braneworld
theory \cite{vdvz,lss}.

On small distances from the central source, $r \ll r_*$, both positive-tension
and negative-tension branes apparently behave similarly reproducing the
Einstein gravity to a high precision with the gravitational constant $1/m^2$,
which is the bare gravitational coupling in the braneworld action
(\ref{action}). However, this expectation is to be verified by refined
calculations in braneworld models with arbitrary sign of brane tension and
without the RS constraint (\ref{lambda-rs}).  In this respect, we should note
that the solution for a spherically symmetric source (the analog of the
Schwarzschild and interior solution in general relativity) largely remains an
open problem in braneworld theory (for recent progress in the DGP model, see
\cite{GI}).

Specification of boundary conditions is of crucial importance for the
construction of a specific braneworld model since they determine the
inhomogeneous gravitational physics on the brane. From the general viewpoint,
we may regard boundary conditions as any conditions restricting the space of
solutions of the brane--bulk system of equations (\ref{bulk}),
(\ref{brane}).\footnote{This can be compared to the view on boundary conditions
in quantum cosmology, which are regarded there as any conditions suitably
restricting the space of the wave functions of the universe. In particular, the
boundary condition may consist in the universe having no boundary \cite{HH}.}
One of the possible proposals in braneworld theory that we put forward in this
paper is to specify the boundary conditions on the visible brane, and we
discussed one of the simplest such specifications, given by Eq.~(\ref{bc}). In
this case, the equations of the theory become completely closed on the brane
and allow for both positive and negative brane tensions preserving all
cosmological properties. Moreover, this proposal might be extended to the
braneworld theory with timelike extra dimension preserving its attractive
features \cite{bouncing} but without inheriting its nonphysical properties (the
tachyonic character of the Kaluza--Klein gravitational modes). It should be
stressed, however, that the homogeneous and isotropic braneworld cosmology is
uniquely described by Eq.~(\ref{solution}) and, apart from possible restriction
on the dark-radiation constant $C$, is not sensitive to the specification of
boundary conditions for system (\ref{bulk}), (\ref{brane}).

The cosmological model under consideration appears to safely pass the existing
constraints on the variation of the gravitational constant from primordial
abundances of light elements synthesized in the big-bang nucleosynthesis (BBN)
and from cosmic microwave background (CMB) anisotropy \cite{UIY}.  The value of
the gravitational constant at the BBN epoch in our model coincides with the
value measured on small scales (\ref{g-obs}), and the effective gravitational
constant (\ref{g-tilde}) or (\ref{g-eff}) that might affect the large-scale
dynamics of the universe responsible for the CMB fluctuations is within the
uncertainties estimated in \cite{UIY}.

Cosmic mimicry in braneworld models is most efficient in the case of parameter
$\alpha \sim 1$, which, according to (\ref{alpha}), implies that the two
spatial scales, namely, the brane length scale given by (\ref{ell}) and the
curvature scale of the bulk are of the same order: $\ell \sim \ell_{\rm warp} =
\sqrt{- 6 / \Lambda_\b}\,$. This coincidence of the orders of magnitude of
completely independent scales can be regarded as some tuning of parameters,
although it is obviously a mild one.

\ack

One of the authors (VS) thanks Bruce Bassett for an interesting conversation.
The authors acknowledge support from the Indo-Ukrainian program of cooperation
in science and technology sponsored by the Department of Science and Technology
of India and Ministry of Education and Science of Ukraine.

\end{document}